\documentclass[aps,prl,twocolumn,groupedaddress]{revtex4}

\usepackage{graphics}
\usepackage{epsfig}
\renewcommand{\vec}[1]{\mbox{\boldmath $#1$}}

\begin{document}

\title{Coherent structures in localised and global pipe turbulence}

\author{Ashley P. Willis}
\email{A.Willis@bris.ac.uk}
\author{Rich R. Kerswell}
\email{R.R.Kerswell@bris.ac.uk}

\affiliation{Department of Mathematics, University of Bristol, University Walk,
   Bristol BS8 1TW, United Kingdom}


\begin{abstract}
  The recent discovery of unstable travelling waves (TWs) in pipe flow
  has been hailed as a significant breakthrough with the hope that
  they populate the turbulent attractor. We confirm the existence of
  coherent states with internal fast and slow streaks commensurate in
  both structure and energy with known TWs using numerical simulations
  in a long pipe. These only occur, however, within less energetic
  regions of (localized) `puff' turbulence at low Reynolds numbers
  ($Re=2000-2400$), and not at all in (homogeneous) `slug' turbulence at
  $Re=2800$.  This strongly suggests that all currently known TWs 
  sit in an intermediate region of phase space between the laminar and
  turbulent states rather than being embedded within the turbulent
  attractor itself.  New coherent fast streak states with strongly
  decelerated cores appear to populate the turbulent attractor
  instead.
\end{abstract}


\maketitle

The transition to turbulence in wall-bounded shear flows is a
classical problem that has challenged physicists for over a century.
Some flows, such as that between differentially-heated parallel plates
or between rotating concentric cylinders, exhibit a smooth progression
to increasingly complicated flows via an initial linear instability.
Plane Couette flow and pipe flow, however, abruptly adopt a turbulent
state.  The problem is further complicated by changes in the
spatio-temporal character of the observed flows at different flow
rates.  Pipe flow exhibits a quasi-stable localised turbulent `puff'
state  as well as a globally turbulent `slug' flow \cite{wygnanski73}.
A finite amplitude disturbance is required to trigger turbulence, the
amplitude of which has been shown to depend critically on its `shape'
\cite{peixinho07}.  Similarly, plane Couette flow exhibits localised
spots of turbulence which may be either short lived transients or
survive to arbitrarily long times \cite{bottin98}. Plane Poiseuille
flow, on the other hand, exhibits a linear instability but at flow
rates well beyond those at which turbulence is typically observed.

\begin{figure}
   \epsfig{figure=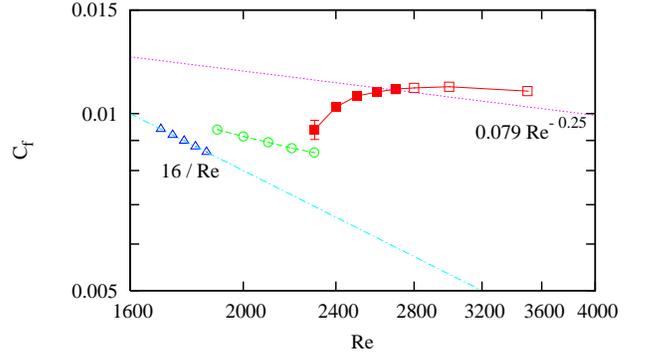, angle=0, width=80mm}
   \caption{ \label{fig:Cf}
      Skin friction coefficient
      $C_f = -\langle \partial_z p \rangle_{r\theta zt}\, D/2\rho
      U^2$ for a $50\,D$ pipe where $\rho$ is the fluid density.
      Triangles - laminar flow; Circles - puffs;
      solid Squares - spatially-inhomogeneous turbulence
      and open Squares - homogeneous slug turbulence (the transition is gradual).  The lower
      straight line corresponds to laminar flow and the upper to the Blasius
      friction law which is initially overshot by the turbulent flow  \cite{priymak04}.
      At $Re=2300$ the flow alternates between a 1 puff state (Circle) and 2 or 3
      puff states (Square with errorbars).
   }
\end{figure}

\begin{figure*}
   \epsfig{figure=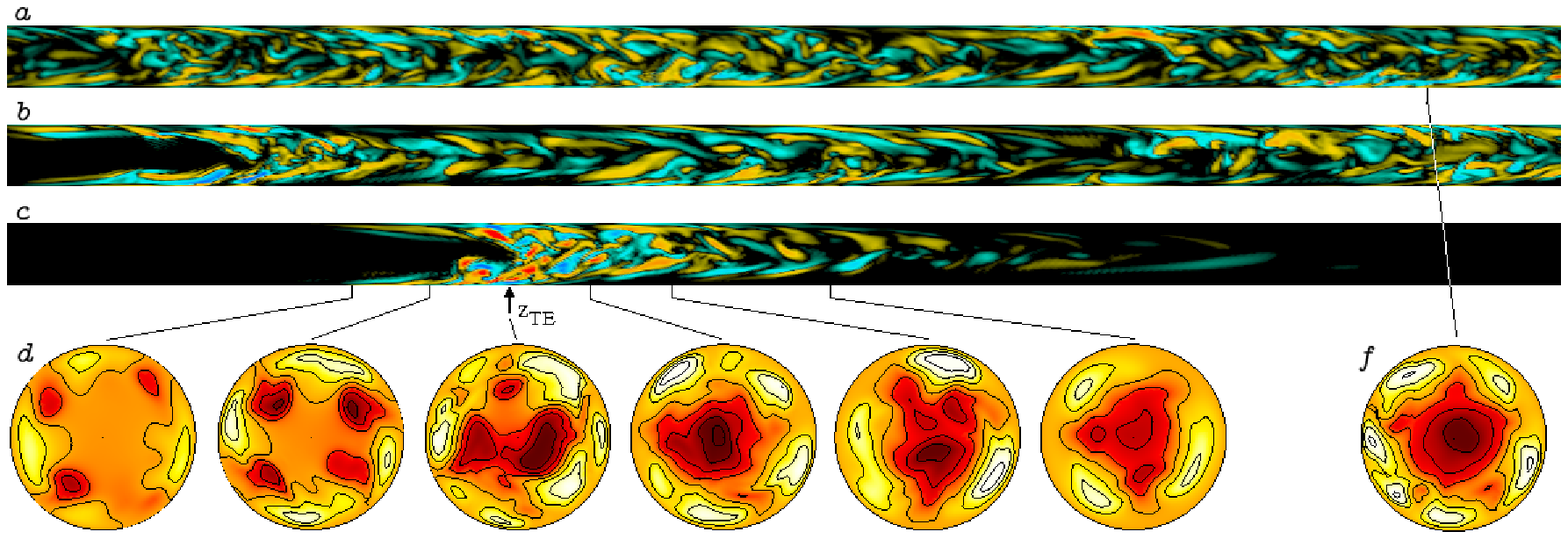, angle=0, width=170mm}\\
   \epsfig{figure=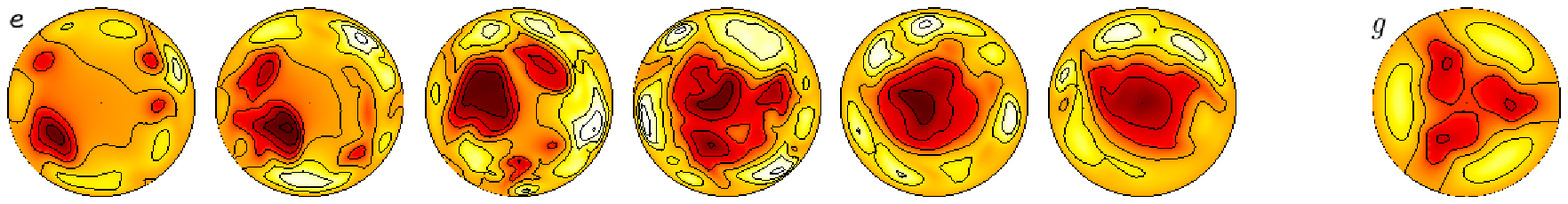, angle=0, width=170mm}
   \caption{ \label{fig:puffs}
      Axial component of vorticity in $(r,z)$-plane,
      $25\,D$ shown of $50\,D$ computational domain;
      ({\it a}) slug turbulence at $Re=2800$; ({\it b}) inhomogeneous turbulence at $Re=2400$
      ({\it c}) puff turbulence at $Re=2000$.
       Cross-sections in $(r,\theta)$ show the axial flow relative
      to the laminar profile with fast streaks (light/white) and slow streaks (dark/red), contour lines
      each $0.2\,U$;
      ({\it d}) $m=4$ and $m=3$ structures seen upstream and downstream of the trailing
      edge $z_{\mathrm{TE}}$ (flow is left to right);
      ({\it e}) sections from a puff where no clear structures are observed;
      ({\it f}) energetic section at $Re=2800$, but resembling $m=5$ structure;
      ({\it g}) exact solution with $3$-fold rotational symmetry.
   }
\end{figure*}

The discovery of exact travelling wave (TW) solutions in wall-bounded
shear flows \cite{nagata90, waleffe98, waleffe01, faisst03, wedin04,
  pringle07} has spurred a flurry of excitement within the community
and prompted the speculation that the states lie inside the turbulent
attractor. Since TWs appear only to have a few unstable directions in
phase space \cite{faisst03,kerswell07}, a turbulent flow trajectory is
imagined to wander between these states, with the probability of the
flow `visiting' a particular TW determined by how unstable it is. The
hope is that turbulent statistics are then predictable from an
appropriately-weighted sum of all the relevant TW properties (see the
reviews \cite{kerswell05, eckhardt07}).

Pipe flow has emerged as the main setting to confirm this picture.
Thus far, numerical results have been confined to unrealistically
short periodic pipes \cite{kerswell07,schneider07}, and while
experimental observations suggest evidence of TWs, important
structural differences remain (compare figs 2E \& F and 4 in
\cite{hof04} and fig 1 in \cite{hof05}).  In this Letter, we use
direct numerical simulations in a sufficiently long pipe to capture
real localised (`puff') and global (`slug') turbulent states, in order
to determine whether TWs populate the turbulent attractor.  The
ability to accurately simulate spatially-inhomogeneous turbulence at
transitional Reynolds numbers proves crucial in revealing the true
position of the TWs in phase space.


The non-dimensionalised governing Navier--Stokes equations for an
incompressible fluid are
\begin{equation}
\partial_t \vec{u}+\vec{u}\cdot\vec{\nabla} \vec{u}+ \vec{\nabla} p
= {1 \over Re} \nabla^2 \vec{u}, \qquad \nabla\cdot\vec{u}=0,
\end{equation}
where the Reynolds number $Re:=UD/\nu$ ($U$ is the mean axial flow
speed, $D$ is the pipe diameter and $\nu$ is the fluid's kinematic
viscosity). These are solved in a pipe of length $50\,D$ across
which periodic boundary conditions are imposed at constant $Re$
(mass flux) in cylindrical coordinates $(r,\theta,z)$ using the
formulation and numerical resolution described in \cite{willis07}.
(At $Re=2800$, the energy of spectral coefficients falls by at least
6 orders of magnitude with spectral order in either $r, \theta$ or
$z$.)

%
%
%
Data was collated from simulations over a range of $Re$ up to $3500$
and over times of greater than $3000\,D/U$; the skin friction is a
useful indicator of the flow response --- see Fig.\ \ref{fig:Cf}. A
localised `puff' structure \cite{wygnanski73} of apparently stable
length $\approx 20\,D$ is observed for $Re \lesssim 2250$. At $2250
\lesssim Re\lesssim 2500$ the puff gradually expands while translating
along the pipe, possibly dividing into multiple puffs, giving rise to
an uneven patch of turbulence in which the turbulent intensity is
spatially inhomogeneous (see Fig.\ \ref{fig:puffs}{\it a-c}). By
$Re \approx 2800$, the turbulent intensity has become much more spatially
homogenized indicating `slug'-like turbulence \cite{wygnanski73}.

\begin{figure}
   \epsfig{figure=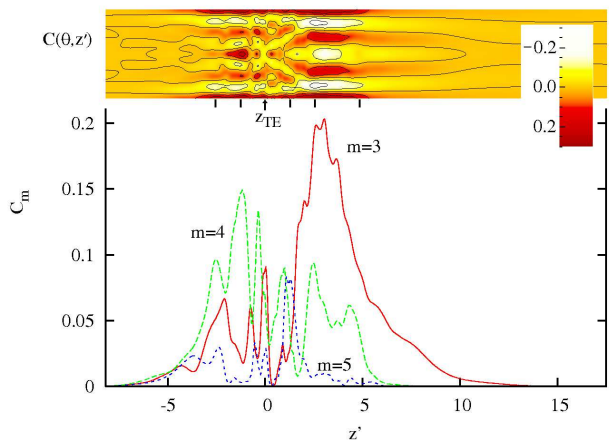, angle=0, width=85mm}
   \caption{ \label{fig:Zmcorr}
      Instantaneous correlations for the puff snapshot of
      Fig.\ \ref{fig:puffs}{\it c}
      at positions relative to the trailing edge $z'=z-z_{\mathrm{TE}}$.
      In the upper plot $\theta$ goes from $0$ to $2\pi$ vertically, contour intervals
      of $0.1$. The marks
      below indicate the positions of the cross-sections of Fig.\ \ref{fig:puffs}{\it d}.
   }
\end{figure}

As a puff is spatially inhomogeneous, the search for coherent
structures was conducted at fixed relative positions, up- and down-
stream, of the puff's steadily translating trailing edge,
$z_{\mathrm{TE}}(t)$, which is itself characterized by a sharp jump in
the streamwise velocity $u_z$ on the pipe axis \footnote{ The position
  of an $\mathrm{e}^{-1}$ drop from the peak to background value in
  $u_z$ on the axis was used to set the location of the trailing edge,
  $z_{\mathrm{TE}}$, where $u_z$ was smoothed over $\pm 1\,D$ to avoid
  jumps when monitoring $z_{\mathrm{TE}}$ caused by the incursion of
  vortices shed upstream. }. The search focused upon the appearance of
fast streaks near the pipe wall, thus correlations in the streamwise
velocity were examined using the function
\begin{equation}
   \label{eq:corr}
   C(\theta,z') = \left.
   \frac{ 2 \, \langle u'_z(\theta+\phi,z')\,u'_z(\phi,z') \rangle_{\phi} }
   { \langle \max_{\phi,z} (u'_z)^2 \rangle_t }
   \right|_{r=0.4D}
\end{equation}
where $\langle\,\cdot\,\rangle_s$ indicates averaging over the
subscripted variable, and $u'_z$ is the deviation from the
time-averaged profile calculated for each $z'=z-z_{\mathrm{TE}}$
position in the puff. The projection function $C_m(z') = 2 \, \langle
C(\theta,z') \cos(m\theta) \rangle_\theta $ was used to extract the
signature of structures of azimuthal wavenumber $m$.  Experience of
examining flow structures indicated that a `good' correlation is
achieved for $C_m(z')$ larger than $0.1$, as indicated by Fig.\ 
\ref{fig:Zmcorr}, which shows the correlation results for the puff
snapshot of Fig.\ \ref{fig:puffs}{\it c}. The magnitude of the
correlation measures are relatively large at the positions indicated
in Fig.\ \ref{fig:Zmcorr} despite not being located at
$z_{\mathrm{TE}}$, where the turbulent intensity is greatest ($u'_z$
is largest) for the puff. Cross-sections in $(r,\theta)$ of the flow
field are shown in Fig.\ \ref{fig:puffs}{\it d} with lines indicating
their position. For comparison purposes, cross-sections for another
puff snapshot are reproduced in Fig.\ \ref{fig:puffs}{\it e} where
$C_m(z')$ is less than $0.1$ for all $z'$. Particularly for the plots
upstream of the trailing edge, the similarity to known TWs is
remarkable where individual slow streaks are also reproduced in the
interior (compare, for example, with Figs.\ 9{\it a} and 13(lower
left) in \cite{wedin04}).

\begin{figure*}
   \epsfig{figure=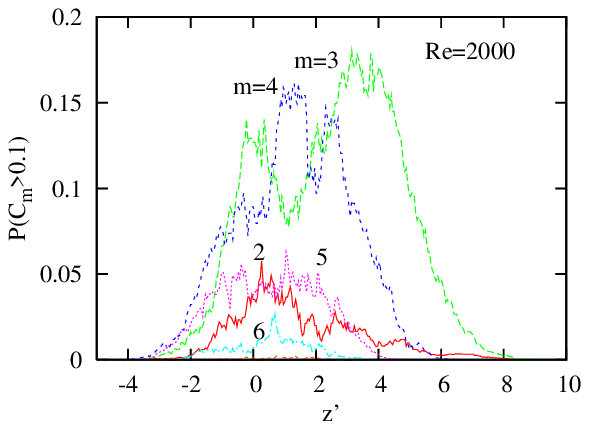, angle=0, width=56mm}
   \epsfig{figure=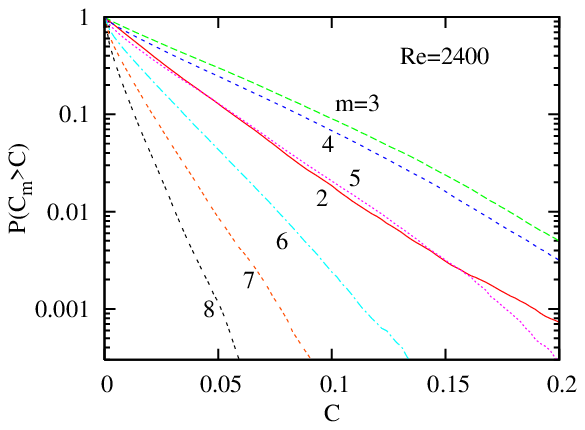, angle=0, width=56mm}
   \epsfig{figure=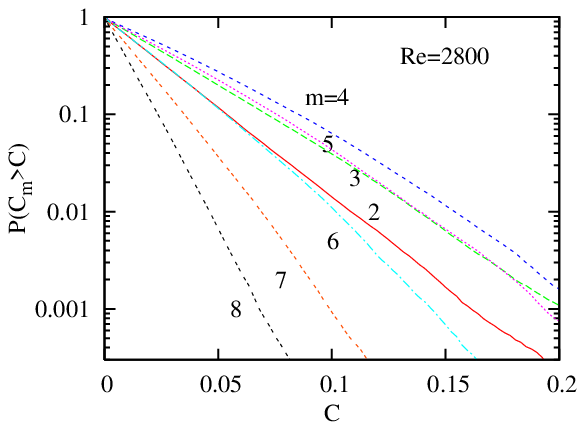, angle=0, width=56mm}
   \caption{ \label{fig:ProbZm}
      Probability at different parts of a puff of a `good' correlation
      $C_m(z') \gtrsim 0.1$ (see Fig.\ \ref{fig:Zmcorr}) at $Re=2000$ (left).
      Probability of a correlation $C_m > C$ at any given point in the flow
      for $Re=2400$ (middle) and $2800$ (right).   }
\end{figure*}
\begin{figure}
   \epsfig{figure=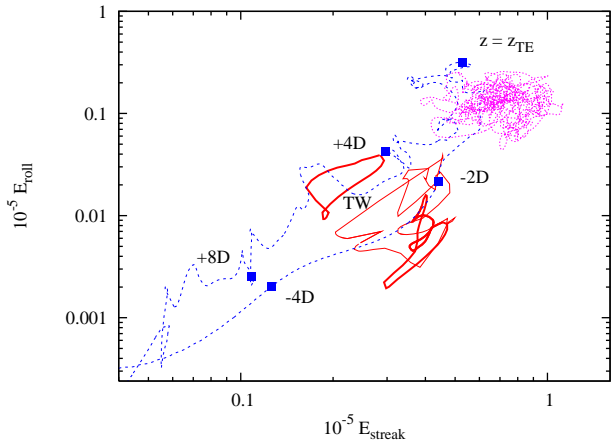, angle=0, width=85mm}
   \caption{ \label{fig:EstErl}
      Roll and streak energies at different parts of the puff of
      Fig.\ \ref{fig:puffs}{\it c} ($Re=2000$).
      Expanding in Fourier modes $m$ in $\theta$, the
      total streak and roll energies are defined as
      $E_{\mathrm{streak}}(z') := Re^2 \pi \sum_{m\ne 0} \int |u'_{m\,z}|^2
      r\,\mathrm{d}r$ and $E_{\mathrm{roll}}(z') := Re^2 \pi \sum_{m\ne 0}
      \int (|u'_{m \,r}|^2 + |u'_{m \,\theta}|^2)\, r\,\mathrm{d}r$ in
      units $\rho \,\nu^2$.
      Here $\vec{u}'=(u'_r,u'_{\theta},u'_z)$ is the deviation from the
      laminar profile, for easier comparison with the TWs.
      Energies for 2-, 3- and 4-fold rotationally-symmetric TWs \cite{kerswell05}
      are shown using lower thick, middle thin and upper thick red solid lines
      respectively (the closed loops are produced by the finite continuum of
      TW axial wavenumbers which can exist at $Re=2000$).  The pink
     short-dotted line `cloud' in the top right hand corner corresponds to a slug
     at $Re=2800$.   }
\end{figure}

The probabilities of finding a correlation greater than $0.1$ at
different parts of the puff are plotted in Fig.\ \ref{fig:ProbZm}.
From around $z'=-D$ to $z'=+5D$ fast streak structures of $m=3$ and
$m=4$ are seen approximately 10--15\% of the time which is in good
agreement with the frequencies observed in
\cite{peixinho06,kerswell07,schneider07}. The appearance of coherent
fast-streaks, however, does not necessarily imply an observation of a
TW. Of the coherent fast streak structures found, those which look
most like TWs (in terms of fast {\em and} slow streaks) are found away
from the most energetic regions in puff turbulence. The disturbance
energy at the trailing edge itself (Fig.\ \ref{fig:EstErl}) is far too
high to be compatible with any known TW at the same $Re$ with the roll
energy, in particular, an order of magnitude too large. Cross-sections
at $z_{\mathrm{TE}}$ also exhibit small-scale structure
uncharacteristic of TWs (see Fig.\ \ref{fig:puffs}{\it g}). Upstream
and downstream, at approximately $z'=-2\,D$ and $z'=+4\,D$, where we
find well formed coherent structures, the magnitudes of streak and
roll energies are both consistent with TWs which all have a
characteristically small roll-to-streak energy ratio.

\begin{figure}
   \begin{center}
      \epsfig{figure=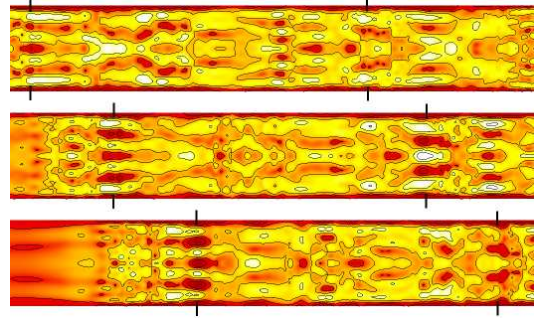, angle=0, width=70mm}
   \end{center}
   \caption{ \label{fig:Zmcorrslug}
      Correlations in inhomogeneous turbulence ($Re=2400$) at time intervals
      of $4\, D/U$, in a fixed window of $25\,D$ of the domain. The middle
      snapshot corresponds to that
      of Fig.\ \ref{fig:puffs}{\it b}.  Same scales as Fig.\ \ref{fig:Zmcorr}.
      The black lines mark positions of waves at the different times
      indicating the translation of the TWs. The phase
      speeds appear $\approx (1 \pm 0.1)U$.
   }
\end{figure}

%
At higher $Re$, the turbulence becomes delocalized as the puffs expand
to invade the whole flow. The strongly turbulent region around the
trailing edge also lengthens to swallow up the weaker relaminarization
zones so that the flow spatially homogenizes.  Correlations can then
be further averaged over the pipe length $C_m := \langle C_m(z)
\rangle_z$ where $u'_z$ is now taken as the deviation from the
full-space-and-time-averaged profile. Fig.\ \ref{fig:ProbZm} shows a
clear trend in which the preferred structures are gradually decreasing
in scale from $m=3$ and $4$ streaks at $Re=2000$ to $m=4$ and $5$
streak structures at $Re=2800$. A typical correlation episode in split
puffs at $Re=2400$ is shown in Fig.\ \ref{fig:Zmcorrslug} using three
snapshots $4\,D/U$ apart. There are two coherent structures
simultaneously present in the $25\,D$ section shown which translate
with phase speeds of $\approx (1\pm 0.1\,)U$. As in the puff, however,
transient signatures of TWs are found to occur in the less energetic
regions of puff turbulence.

At larger Reynolds numbers, $Re \gtrsim 2800$, the turbulent intensity
in slug turbulence is uniformly high everywhere, being comparable to
that at the trailing edge of a puff (in units of $\rho\,\nu^2$ at each
cross-section), and is never as low as that of the TWs (see Fig.
\ref{fig:EstErl}). Although fast streak structures are still observed,
they are of too high energy to be associated with known TWs.  The
cross-section shown in Fig.\ \ref{fig:puffs}{\it f} ($C_5 >0.1$) from
a slug is typical, where a large and strongly retarded central core
dominates, with only narrow fingers extending towards the gaps between
the intense fast streaks at the wall \footnote{Plotting the flow
  relative to a different profile may change Fig.\ \ref{fig:puffs}{\it f}\, 
  slightly, and may reduce the relative streak energy of the perturbation, but
  cannot resolve the disparity in roll energies.}. This retarded core
feature of the coherent structures found at $Re \gtrsim 2800$ can now
be appreciated as a significant problem in previous qualitative
comparisons between slug cross-sections and known TWs
\cite{hof04,hof05}. It is more likely that these coherent structures
instead point to the existence of other types of exact, more highly
nonlinear solutions.

%
%

Our results suggest that the known TWs populate an intermediate region
of phase space between the laminar and fully turbulent phases, rather
than the turbulent part of phase space itself. The dynamical
importance of the TWs is therefore in the transition-to-turbulence
process where the fate of an initial disturbance is of concern rather
than in characterizing the established turbulent state. The puff
provides a simple illustrative example of this picture of TWs sitting
between the laminar and turbulent states. At $Re=2000$, a puff travels
at only $\approx 90\%$ of the bulk velocity so, on average, fluid
passes through it.  Far upstream ($t \rightarrow -\infty$), the fluid
`trajectory' starts at the origin (laminar state), passes through the
TW-region of phase space as $t$ increases, to reach the fully
turbulent region near the trailing edge. On leaving here as time
increases, it passes back through the TW region to the origin as it
relaminarises far downstream ($t \rightarrow +\infty$). Consequently,
TWs are only visited just up- and down- stream of the trailing edge:
Fig \ref{fig:EstErl} is a good 2-D representation of this process
(where $z$ plays the role of $t$ and a phase space norm based on the
streak and roll energies is implied).

Our results confirm emerging evidence that lower branch TWs lie
strictly between the laminar and turbulent states in phase space
\cite{itano01,kerswell07,wang07,duguet08}. However, the fact that
upper branches of the known TWs also have too low energy to be
associated with the turbulent part of phase space is a surprise. It is
quite plausible that the coherent structures observed so far for
$Re\gtrsim 2800$ and characterised by an outer ring of fast streaks
together with a strongly decelerated core represent a more-energetic
branch of TWs which {\em is} embedded in the turbulent attractor.  The
fact that their roll-to-streak energy ratio is so much larger than for
currently known TWs suggests that they may have a different sustaining
mechanism.

\begin{acknowledgments}
  We thank Jorge Peixinho and Tom Mullin for many valuable
  discussions, and particularly Tom for challenging us to produce
  figure 2. This research was funded by the EPSRC under grant
  GR/S76144/01.
\end{acknowledgments}

\bibliography{paper_resub}

\end{document}